\documentclass[aps,showpacs,pre,twocolumn,floatfix]{revtex4}

\usepackage[english]{babel}
\usepackage{subfigure}
\usepackage{latexsym}
\usepackage{graphicx}
\usepackage{epsfig}

\begin{document}

\title{Numerical Methods for Stochastic Differential Equations}
\author{Joshua Wilkie}
\affiliation{Department of Chemistry, Simon Fraser University, Burnaby,
             British Columbia V5A 1S6, Canada}

\begin{abstract}
Stochastic differential equations (sdes) play an important role in physics but existing numerical methods for solving such equations are of low accuracy and poor stability. A general strategy for developing accurate and efficient schemes for solving stochastic equations in outlined here. High order numerical methods are developed for integration of stochastic differential equations with strong solutions. We demonstrate the accuracy of the resulting integration schemes by computing the errors in approximate solutions for sdes which have known exact solutions.
\end{abstract}

\maketitle


Stochastic differential equations ({\bf sdes}) have a long history in physics\cite{CWG} and play an important role in many other areas of science, engineering and finance\cite{CWG,Hase,Platen}. Recently a number of computational techniques have been developed in which high dimensional deterministic equations are decomposed into lower dimensional stochastic equations. Gisin and Percival\cite{GP}, for example, reduced a deterministic master equation for the density matrix into stochastic equations for a wavefunction. Similar approaches are being used to solve the quantum many-body problem for bosons\cite{CCD}, fermions\cite{JC} and vibrations\cite{Wilk}. These latter methods give rise to large sets of coupled sdes which require fast and efficient numerical integration schemes. Unfortunately, and in spite of their widespread use, the available numerical techniques\cite{Platen} for solving such equations are far less accurate than comparable methods for solution of ordinary differential equations ({\bf odes}).

In this manuscript we show how classical methods for solving odes, such as Runge-Kutta, can be adapted for the solution of a class of sdes which should include many of the equations which arise in physical problems. 

Consider a finite set of sdes,
\begin{eqnarray}
dX^j_t=a^j({\bf X}_t,t)~dt+\sum_{k=1}^mb^j_k({\bf X}_t,t)~dW^k_t,
\label{sdes}
\end{eqnarray}
represented in It\^{o}\cite{CWG,Hase,Platen} form, where $j=1,\dots,n$. Here ${\bf X}_t=(X^1_t,\dots,X^n_t)$ and the $dW^k_t$ are independent and normally distributed stochastic differentials with zero mean and variance $dt$ (i.e. sampled $N(0,dt)$). The stochastic variables $W^k_t$ are Wiener processes. Now assume that the coefficients $a^j$ and $b^j_k$ have regularity properties which guarantee strong solutions, i.e. that $X^j_t$ are some fixed functions of the Wiener processes, and that they are differentiable to high order. [Sufficient conditions for strong solutions are discussed in Ref. \cite{Platen}.] We may then view the solutions of (\ref{sdes}) as functions $X^j_t=X_j(t,W^1_t,\dots, W^m_t)$ of time and the Wiener processes. The solutions can therefore be expanded in Taylor series. Keeping terms of order $dt$ or less then gives
\begin{eqnarray}
X^j_{t+dt}&=&X^j_t+\frac{\partial X^j_t}{\partial t}~dt+\sum_{k=1}^m\frac{\partial X^j_t}{\partial W^k_t}~dW^k_t\nonumber \\
&+&\frac{1}{2}\sum_{k,l=1}^m\frac{\partial^2 X^j_t}{\partial W^k_t\partial W^l_t}~dW^k_tdW^l_t.\label{sdes2}
\end{eqnarray}
In a mean square sense the product of {\em differentials} $dW^k_tdW^l_t$ is equivalent to $\delta_{k,l}dt$ in the It\^{o}\cite{CWG,Hase,Platen} formulation of stochastic calculus. Making this replacement then yields
\begin{eqnarray}
dX^j_{t+dt}&=&X^j_{t+dt}-X^j_t=[\frac{\partial X^j_t}{\partial t}+\frac{1}{2}\sum_{k=1}^m\frac{\partial^2 X^j_t}{\partial W^{k2}_t}]~dt\nonumber \\
&+&\sum_{k=1}^m\frac{\partial X^j_t}{\partial W^k_t}~dW^k_t
\end{eqnarray}
which when compared to (\ref{sdes}) allows us to identify the first derivatives
\begin{eqnarray}
\frac{\partial X^j_t}{\partial W^k_t}&=&b^j_k({\bf X}_t,t) \\
\frac{\partial X^j_t}{\partial t}&=&a^j({\bf X}_t,t)-\frac{1}{2}\sum_{k=1}^m\frac{\partial^2 X^j_t}{\partial W^{k2}_t}\nonumber \\
&=&a^j({\bf X}_t,t)-\frac{1}{2}\sum_{k=1}^m\sum_{i=1}^nb^i_k({\bf X}_t,t)\frac{\partial b^j_k({\bf X}_t,t)}{\partial X_t^i}.
\end{eqnarray}
Now that these first order derivatives are expressed in terms of $a^j$ and $b^j_k$, higher order derivatives can be computed. Thus a Taylor expansion of the solutions 
\begin{eqnarray}
X^j_{t+\Delta t}&=&X^j_t+\frac{\partial X^j_t}{\partial t}\Delta t+\sum_{k=1}^m\frac{\partial X^j_t}{\partial W^k_t}~\Delta W^k_t\nonumber \\
&+&\frac{1}{2}\sum_{k,l=1}^m\frac{\partial^2 X^j_t}{\partial W^k_t\partial W^l_t}\Delta W^k_t\Delta W^l_t+\dots
\end{eqnarray}
can be obtained for finite displacements $\Delta t$ and $\Delta W^k_t$. This Taylor expansion can then be employed to develop Runge-Kutta algorithms and other integration schemes.

We illustrate the use of this approach by developing a Runge-Kutta method for sdes which is closely related to the classical Runge-Kutta scheme for odes. For given displacements $\Delta t$ and $\Delta W^k_t$ define
\begin{eqnarray}
f_j({\bf X}_t,t)&=&\frac{\partial X^j_t}{\partial t}\Delta t+\sum_{k=1}^m
\frac{\partial X^j_t}{\partial W^k_t}\Delta W^k_t\nonumber \\
&=&[a^j({\bf X}_t,t)-\frac{1}{2}\sum_{k=1}^m\sum_{i=1}^nb^i_k({\bf X}_t,t)\frac{\partial b^j_k({\bf X}_t,t)}{\partial X_t^i}]\Delta t\nonumber \\
&+&\sum_{k=1}^mb^j_k({\bf X}_t,t) \Delta W^k_t
\end{eqnarray}
and consider the following four stage approximation
\begin{eqnarray}
K_j^1&=&f_j({\bf X}_{t_i},t_i)\nonumber \\
K_j^2&=&f_j({\bf X}_{t_i}+\frac{1}{2}{\bf K}^1,t_i+\frac{1}{2}\Delta t)\nonumber \\
K_j^3&=&f_j({\bf X}_{t_i}+\frac{1}{2}{\bf K}^2,t_i+\frac{1}{2}\Delta t)\nonumber \\
K_j^4&=&f_j({\bf X}_{t_i}+{\bf K}^3,t_{i+1})\nonumber\\
{\bf X}_{t_{i+1}}&=&{\bf X}_{t_i}+\frac{1}{6}({\bf K}^1+2{\bf K}^2+2{\bf K}^3+{\bf K}^4)
\label{rk4}
\end{eqnarray}
where $t_i$ is the initial time and $t_{i+1}=t_i+\Delta t$. Taylor expansion of this scheme shows that ${\bf X}_{t_{i+1}}$ differs from the exact solution by terms of order higher than $\Delta t^2$ (i.e. terms of higher order than $\Delta t^2$, $\Delta t (\Delta W^k_t)^2$, $(\Delta W^k_t)^4$, $(\Delta W^k_t)^2(\Delta W^l_t)^2$, and $(\Delta W^k_t)^2\Delta W^l_t\Delta W^i_t$). Thus, this stochastic Runge-Kutta algorithm plays a role very similar to its classical counterpart except that its order is reduced from four to two. Generalizations to higher order Runge-Kutta schemes are straightforward, and we will employ one such scheme in example calculations, but details will not be presented here.

While this approach is not completely general, since it will fail for sdes with weak solutions or non-differentiable $a^j$ and $b^j_k$, it should be applicable to a wide range of problems. It can for example be used to solve every one of the equations with known solutions tabulated in section 4.4 of Ref. \cite{Platen}. To illustrate the accuracy of the method and its improvement over other known techniques for solving sdes we now consider a number of these examples. We compare known exact solutions with numerical solutions obtained using the Euler-Maruyama scheme\cite{EM}, a derivative free version of the Milstein scheme due to Kloeden and Platen\cite{Mils}, the classical Runge-Kutta scheme (\ref{rk4}), and another Runge-Kutta scheme obtained in the manner outlined above from an eighth order twelve step method for odes due to Hairer and Wanner\cite{Hair} (this reproduces the stochastic Taylor expansion up to and including terms of order $\Delta t^4$). Stochastic differentials were sampled using the routines gasdev and ran2\cite{NR}.
\begin{figure}
\caption{$\log_{10}|X_t-X_t^{approximate}|$ vs time $t$ for Eq. (\ref{EG1})}
\includegraphics[width=3.4in]{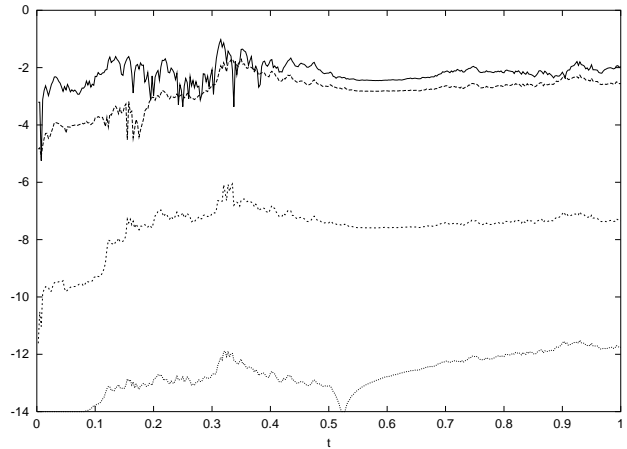}
\label{Fig1}
\end{figure}

As a first test of these methods consider an autonomous nonlinear scalar equation
\begin{equation}
dX_t = (1+X_t)(1+X_t^2) dt + (1+X_t^2) dW_t \label{EG1}
\end{equation}
with just one Wiener process. In this example and in all subsequent examples we assume all Wiener processes are initially zero. The exact solution to this equation is\cite{Platen}
\begin{equation}
X_t=\tan(t+W_t+\arctan(X_0))
\end{equation}
as can be readily verified using It\^{o}\cite{CWG,Hase,Platen} calculus. In Fig. 1 we plot the error $\log_{10}|X_t-X_t^{approximate}|$ vs time computed with a time step of $2.5\times 10^{-5}$ for a single stochastic trajectory with initial condition $X_0=1$ for the four different approximation schemes. The Milstein scheme (long-dashed curve) shows some improvement over the primitive Euler-Maruyama method (solid curve) but the order two Runge-Kutta scheme (short-dashed curve) and order four Runge-Kutta scheme (dotted curve) perform very much better.
\begin{figure}
\caption{$\log_{10}|X_t-X_t^{approximate}|$ vs time $t$ for Eq. (\ref{EG2})}
\includegraphics[width=3.4in]{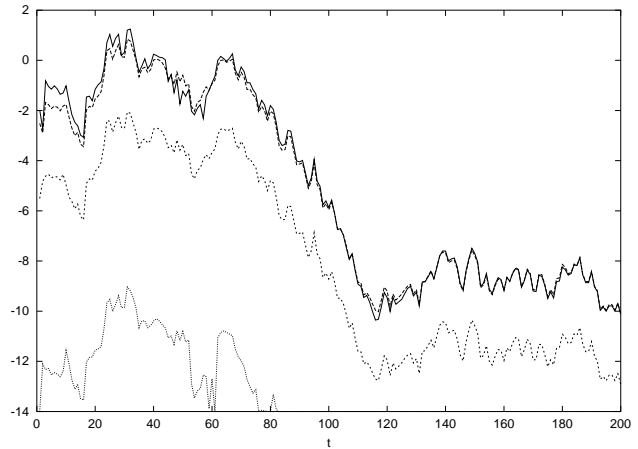}
\label{Fig2}
\end{figure}

The second example equation, also from Ref. \cite{Platen}, is an autonomous linear scalar equation in two Wiener processes
\begin{equation}
dX_t = a_0 X_t dt + b_1 X_t dW^1_t + b_2 X_t dW^2_t\label{EG2}
\end{equation}
which has an exact solution 
\begin{equation}
X_t=X_0\exp\{[a_0-\frac{1}{2}(b_1^2+b_2^2)]t+b_1 W_t^1+b_2W_t^2\}.
\end{equation}
The logarithm base ten of the error for the different schemes, calculated for initial condition $X_0=1$ and time step .01, is plotted in Fig. 2. Here the Milstein scheme (long-dashed curve) performs no better than the Euler-Maruyama method (solid curve) but again the order two Runge-Kutta scheme (short-dashed curve) and order four Runge-Kutta scheme (dotted curve) show greatly improved accuracy. [Note that the apparent improvement in performance of all schemes at long time is a result of the fact that the solution decays to zero.]
\begin{figure}
\caption{$\log_{10}|X_t^1-X_t^{1~approximate}|$ vs time $t$ for Eq. (\ref{EG3})}
\includegraphics[width=3.4in]{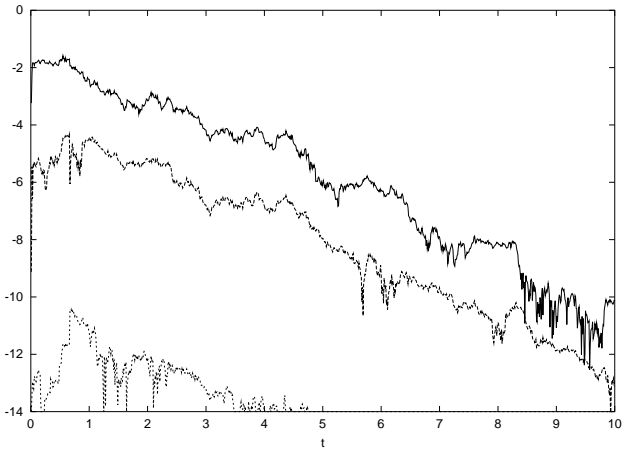}
\label{Fig3}
\end{figure}

Example 3 is a set of two coupled linear autonomous sdes 
\begin{eqnarray}
dX_t^1&=&-\frac{3}{2} X_t^1 dt +X_t^1 dW_t^1-X_t^1dW_t^2-X_t^2dW_t^3\nonumber \\
dX_t^2&=&-\frac{3}{2} X_t^2 dt +X_t^2 dW_t^1-X_t^2dW_t^2+X_t^1dW_t^3\label{EG3}
\end{eqnarray}
with three Wiener processes. Here the solutions are 
\begin{eqnarray}
X_t^1&=&\exp\{-2t+W_t^1-W_t^2\}\cos W_t^3\nonumber\\
X_t^2&=&\exp\{-2t+W_t^1-W_t^2\}\sin W_t^3.
\end{eqnarray}
Numerical solutions were calculated with a time step of .01 and errors in $X_t^1$ are represented in Fig. 3. The order two Runge-Kutta scheme (long-dashed curve) and order four Runge-Kutta scheme (short-dashed curve) show improvement over the Milstein scheme (solid curve). Similar results were obtained for $X_t^2$.
\begin{figure}
\caption{$\log_{10}|X_t-X_t^{approximate}|$ vs time $t$ for Eq. (\ref{EG4})}
\includegraphics[width=3.4in]{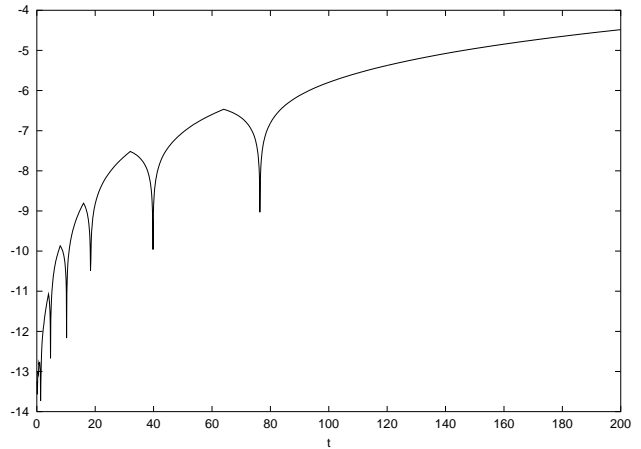}
\label{Fig4}
\end{figure}

The examples we have considered so far have not had explicitly time dependent $a^j$ and $b^j_k$. Example 4 is a scalar non-autonomous sde
\begin{equation}
dX_t=[\frac{2}{1+t}X_t+\frac{1}{2}(1+t)^2]dt+\frac{1}{2}(1+t)^2 dW_t\label{EG4}
\end{equation}
with known solution\cite{Platen}
\begin{equation}
X_t=\left(\frac{1+t}{1+t_0}\right)^2X_0+\frac{1}{2}(1+t)^2(W_t+t-t_0).
\end{equation}
Numerical solutions were calculated using the order two Runge-Kutta scheme and a time step of .001, $t_0=0$ and $X_0=1$.
The error is represented in Fig. 4. As in previous examples a high accuracy is achieved in spite of the rapid growth of the solution. The comparative smoothness of the error curve reflects the fact the the deterministic part of the solution dominates.
\begin{figure}
\caption{$\log_{10}|X_t-X_t^{approximate}|$ vs time $t$ for Eq. (\ref{EG5})}
\includegraphics[width=3.4in]{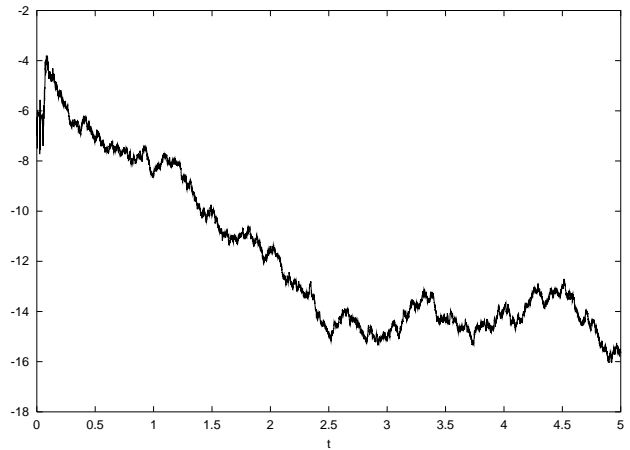}
\label{Fig5}
\end{figure}

We now consider an example for which an exact solution is known but which is expressed in terms a stochastic integral. Consider the stochastic Ginzburg-Landau equation
\begin{equation}
dX_t=[-X_t^3+(\alpha+\frac{1}{2}\sigma^2)X_t]dt+\sigma X_t dW_t\label{EG5}
\end{equation}
with solution\cite{Platen}
\begin{equation}
X_t=X_0\frac{\exp\{\alpha t+\sigma W_t\}}{\sqrt{1+2X_0^2\int_0^t\exp\{2\alpha s+2\sigma W_s\}ds}}.
\end{equation}
We chose $\alpha=.01$, $\sigma=4$, $X_0=1$ and $dt=5\times 10^{-6}$. The stochastic integral was computed using a Riemann sum with the same time step. Error in the solution calculated with the order two Runge-Kutta scheme is plotted in Fig. 5. Good accuracy is again obtained.
\begin{figure}
\caption{$\log_{10}|X_t-X_t^{approximate}|$ vs time $t$ for Eq. (\ref{EG6})}
\includegraphics[width=3.4in]{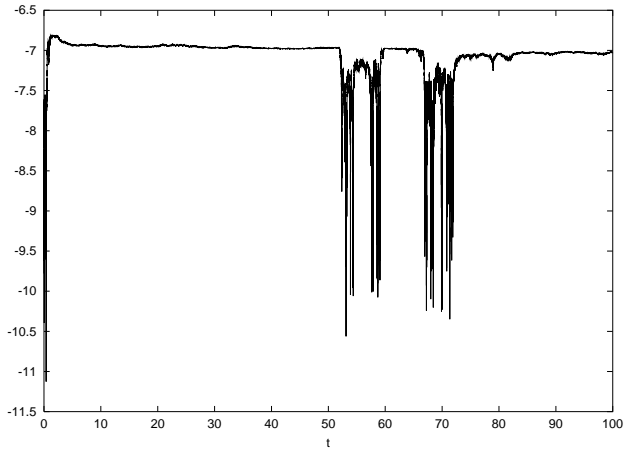}
\label{Fig6}
\end{figure}

Finally, we consider an example in which the exact solution is expressed in terms of a It\^{o}\cite{CWG,Hase,Platen} stochastic integral. Consider the sde
\begin{equation}
dX_t=-\tanh X_t (a+\frac{1}{2}b^2{\rm sech} ^2X_t)dt+b{\rm sech} X_t dW_t\label{EG6}
\end{equation}
with exact solution\cite{Platen}
\begin{equation}
X_t={\rm arcsinh} \left(e^{-at}\sinh X_0+e^{-at}\int_0^t e^{as}dW_s\right).
\end{equation}
\begin{figure}
\caption{$\log_{10}|n_t-n_t^{approximate}|$ vs time $t$ for Eq. (\ref{EG7})}
\includegraphics[width=3.4in]{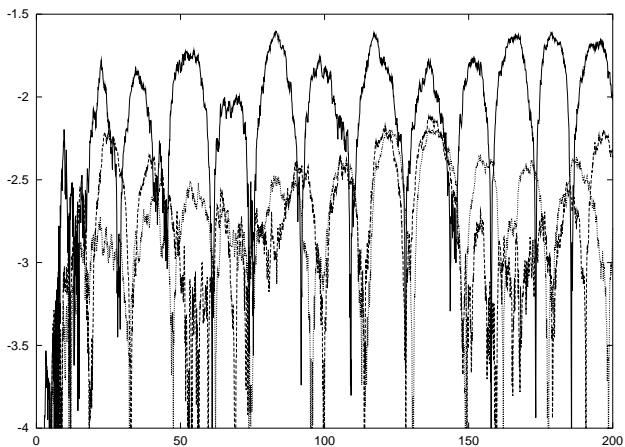}
\label{Fig7}
\end{figure}
We set $a=.02$, $b=1$, $X_0=1$ and $dt=1\times 10^{-5}$. The stochastic integral in the exact solution was calculated using the It\^{o}\cite{CWG,Hase,Platen} integral formula with the same time step. The error in the solution calculated with the order two Runge-Kutta scheme is plotted in Fig. 6. As in all previous cases considered the accuracy is very good.

Thus, the approach to solving sdes advocated here works very well for the wide range of examples we have considered. The order 4 Runge-Kutta method is clearly much more accurate than the order 2 Runge-Kutta scheme. It also has an embedded lower order Runge-Kutta scheme which can be employed 
to obtain an error estimate suitable for stepsize control\cite{Hair}. Hence is should be possible to use 
variable stepsizes to ensure the accuracy of the solution. This sort of implementation is essential for solving equations which do not have known exact solutions. The only subtlety in developing such a method is ensuring that the correct Wiener path is maintained even when a step must be rejected. This is achieved\cite{GL} by dividing the rejected differentials $dt$ and $dW^k_t$ in two segments; $dt/2$ and $dW^k_t/2-y$ followed by $dt/2$ and $dW^k_t/2+y$ where $y$ is sampled $N(0,dt/2)$. To illustrate the accuracy of the resulting variable stepsize algorithm we solve the Gisin-Percival\cite{GP} stochastic wave equation for the nonlinear absorber (Eq. 4.2 of Ref. \cite{GP})
\begin{eqnarray}
d|\psi\rangle &=&.1(a^{\dag}-a)|\psi\rangle dt+(2\overline{a^{\dag 2}}a^2-a^{\dag 2}a^2-\overline{a^{\dag 2}}~\overline{a^2})|\psi\rangle dt\nonumber \\
&+&\sqrt{2}(a^2-\overline{a^2}) |\psi\rangle dW_t\label{EG7}
\end{eqnarray}
with initial state $|\psi(0)>=|0\rangle$. In Fig. 7 we plot the error in mean occupation number $n_t=M[\langle \psi|a^{\dag}a|\psi\rangle ]$ vs time (Fig. 5 of Ref. \cite{GP}) where $M[\cdot ]$ denotes an average over stochastic realisations. 1000, 10000, and 20000 trajectories were used to calculate the solid curve, dashed curve and dotted curve, respectively. Convergence to the exact result is good. 

The author acknowledges the support of the Natural Sciences and Engineering Research Council of Canada.

\end{document}